\documentclass[12pt,onecolumn,showpacs,amssymb,aps,nofootinbib,floatfix]{revtex4-1}
\usepackage{epsfig}
\newcommand{\ave}[1]{\left\langle #1 \right\rangle}

\newcommand{\eqcomma}{\phantom{A},\phantom{A}}

\begin{document}

\title{Indications of a non-trivial vacuum in the effective theory of perfect fluids}
\author{Tommy Burch$\phantom{A}^a$, Giorgio Torrieri$\phantom{A}^b$}
\affiliation{ $\phantom{A}^a$ D-93053 Regensburg, Germany \\ 
$\phantom{A}^b$IFGW, State University of Campinas,Campinas, Sao Paulo, Brazil }
\collaboration{LSS Collaboration}

\begin{abstract}
  Using lattice field theory techniques, we investigate the vacuum structure of the field theory corresponding to 
  perfect fluid dynamics in the Lagrangian prescription. 
  We find intriguing, but inconclusive evidence, that the vacuum of such a 
  theory is non-trivial, casting doubts on whether the gradient expansion can
  provide a good effective field theory for this type of system. 
  The non-trivial vacuum looks like a ``turbulent'' state where some of the 
  entropy is carried by macroscopic degrees of freedom. 
  We describe further steps to strengthen or falsify this evidence.
\end{abstract}

\maketitle

\section{Introduction}

A recent topic of very active interest is to rewrite hydrodynamics as an 
effective field theory \cite{hydro0,hydro1,DerivExp,ritz1}, with the fields 
representing the Lagrangian coordinates of the fluid's volume elements. This 
picture allows the use of well-tested effective field theory techniques to 
investigate fluids in the vanishing viscosity limit, a limit where the very 
definition of hydrodynamics is currently ambiguous \cite{daniel,4pi}. 
An advantage of the field theory approach is that thermodynamic concepts 
like fluid isotropy and entropy conservation can be represented as symmetries.

Phenomenologically, such a theory can be applied to a wide variety of 
settings, from superfluid helium to cosmology \cite{labun} to quark-gluon plasma\cite{hydroplasma}.  From a 
theoretical point of view, it allows us to access a hitherto unexplored region 
\cite{mooresound,ritz2,gthydro}:  One where the mean free path is small enough to 
neglect all dissipative effects but where the microscopic number of particles 
is not ``large'', so microscopic decorrelation (molecular chaos, or, 
equivalently the large $N_c$ limit in an AdS/CFT \cite{4pi} setting) does not 
apply.  Hence,  {\em thermal} fluctuations excite 
hydrodynamic degrees of freedom which subsequently evolve non-linearly: 
When viscosity is so low that ``typical'' sound waves, of frequency $\sim T$ 
and amplitude comparable to a thermal fluctuation, 
$\Delta \rho/\ave{\rho} \sim C_V/T^3$ (where $\rho$ is the energy density, 
$C_V$ the heat capacity and $T$ the temperature), survive for a time much 
larger than the thermal scale, $\sim 1/T$, Kubo's formula needs to be 
renormalized to account for the energy-momentum carried by the sound waves. 
This is a reasonable physical interpretation of applying ``quantum'' concepts, 
such as the definition of observables in terms of functional integrals, to 
something so quintessentially classical as a ``perfect fluid'', 
and might be used to demonstrate that the existence of a quantum limit on 
viscosity is plausible from hydrodynamic arguments alone \cite{gthydro}.

Works such as \cite{DerivExp,cosets,framids} have established a consistent set 
of techniques of generating EFT terms in increasing order of derivatives 
respecting the fundamental symmetries of fluid mechanics, with recent 
dissipative applications \cite{diss1,diss2} becoming possible.  However, it 
is well known that for an EFT expansion to be complete, the right {\em vacuum} 
of the theory has to be known and expanded around.  The presence of 
turbulence in {\em classical} fluid dynamics suggests that vacuum choice 
around the hydrostatic limit, employed in  \cite{DerivExp,cosets,framids}, is 
not necessarily justified, an assumption also discussed in \cite{hydro1}.  

\begin{figure}[h]
\begin{center}
\includegraphics*[clip,width=12.6cm]{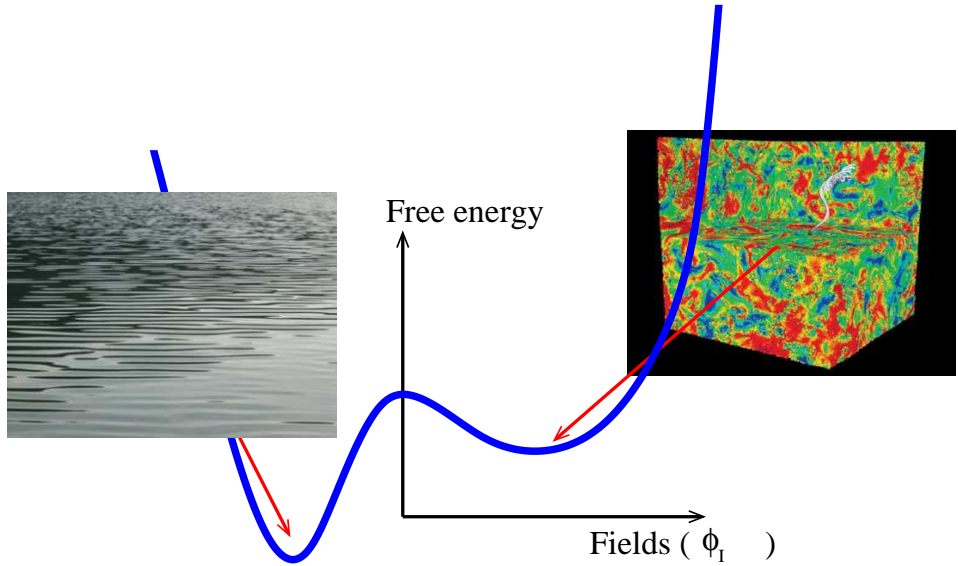}
\caption{\label{vacuum} A schematic illustration of a possible vacuum 
  structure of the theory, in terms of a free energy including both 
  microscopic degrees of freedom and macroscopic collective excitations.  The 
  left side is the "hydrostatic limit", where expansions such as 
  \cite{DerivExp} lead to a well-behaved effective theory.   The right side is 
  a possibly turbulent vacuum where a fraction of the entropy goes into 
  microscopic degrees of freedom.   In this figure the free energy corresponds to the coexistence phase of a first order phase transition, but of course other configurations between the two vacua (second order, cross-over, critical points and so on) are equally possible\\
Image on the right from SCIDA \cite{scida}} 
\end{center}
\end{figure}

Figure \ref{vacuum} illustrates how this could happen:  While the microscopic 
degrees are traced over, they give a contribution to the free energy, 
dimensionally set at $T_0^4$, where $T_0$ is the cutoff scale for microscopic 
degrees of freedom, which needs to be dominant for the hydrostatic limit to apply. 
If the heat capacity at constant pressure of the system is high enough w.r.t. the compressibility (note that the planar limit explicitly excludes this, since there the heat capacity trivially vanishes)
thermal fluctuations will seed waves and vortices which do not go away but will interact non-linearly,
modifying the hydrostatic vacuum.
In terms of fundamental statistical mechanics, this dynamics can be explained by the entropy
carried by the microscopic DoFs (strictly determined through the temperature $T_0$ and the equation of
state) vs the macroscopic DoFs (determined by the attice spacing and the functional integral of the fields).
When the latter becomes a sizeable fraction w.r.t. the former,  part of the microscopic 
entropy will be shared, under conditions of equilibrium, with macroscopic excitations.  If 
this configuration defines a minimum in the free energy, the hydrostatic 
vacuum is unstable and a non-trivial phase diagram results.

Since the presence of turbulent flow destroys the symmetries of the hydrostatic limit (at least compressibility, but in principle homogeneity and even isotropy), the dynamics described above, if it exists, must give rise to a phase transition.
Figure \ref{vacuum} assumes a first-order phase transition, where the two minima 
coexist at a given temperature.   Second-order phase transitions, critical 
points and cross-overs are also possibilities, set by the fluid-dynamical free 
energy in the standard way \cite{landauphase}.

The purpose of this work is to test this assumption using Lattice Monte Carlo 
simulations, the only known way to examine a quantum or statistical theory 
independently of its perturbative structure, continuing the preliminary 
analysis shown in \cite{proc}.

We would like to emphasize, in case of possible misunderstandings, that the
calculations done here are {\em not} real-time simulations of classical
solutions such as \cite{chesler}, but a simulation of a static ``vacuum''
whose partition function is evaluated by a functional integral, analogous to
lattice QCD simulations (and in fact using the same numerical techniques
\cite{Creutz:1984mg,Rothe:1992nt}), 
appropriate to describe, beyond any linearization or perturbative expansion, 
both quantum/thermal fluctuations and their response.    Our results are 
therefore not an evolution of the theory from initial conditions, but a 
description of the ``ground state'' of the theory which, together with the 
Lagrangian, defines the equations of motion.    If this ground state is the 
hydrostatic limit (a finite temperature static fluid), hydrodynamics will 
arise as an effective theory with the gradient defining the coarse-graining 
scale.   If it is not, just as in usual QFTs, vacuum effects will introduce 
additional correlations which will not be visible to any order in the gradient 
expansion (a good analogy here is the role of instantons in QCD \cite{anomalies}).

\section{The theory}

\subsection{Lagrangian description of hydrodynamics}

Three-dimensional ideal hydrodynamics with no chemical potential (any 
``particles'' are balanced by ``antiparticles'', so the net density of each 
conserved charge vanishes in each volume element) can, in the Lagrangian 
picture, be described \cite{hydro0,hydro1} in terms of three fields 
$\phi^{I=1,2,3}$, which physically correspond to the $x,y,z$ coordinates of 
the comoving frame (Lagrangian picture) with respect to the lab frame (in 
which the Eulerian picture is defined). 
Note that unlike scalar fields used in quantum field theory, $\phi^I$ have 
dimensions of one unit of spacetime.

The choice of $\phi^{I=1,2,3}$ is of course not unique, as a perfect fluid is 
homogeneus, and in its comoving frame, invariant under rotations and 
rescalings. 
This symmetry restricts the Lagrangian to the form \cite{hydro0,hydro1} 
\begin{equation}
  \label{lfluid}
  L = F(B) = T_0^4 \tilde F \left( \mathrm{det}\left( B_{IJ} \right)\right) 
  \eqcomma B_{IJ} =   \partial^\mu \phi^I \partial_\mu \phi^J \; . 
\end{equation}
The function $F(B)$ is left arbitrary, as it corresponds to different equations of state for the fluid.
We note that the symmetry automatically determines the interactions, and the coupling constant of the theory, which therefore does not need to be fixed on the lattice.      We also note that any excitations
of the theory can actually be regarded as goldstone modes of one of the broken symmetries included in this Lagrangian \cite{hydro1}.   This makes it imperative to distinguish dynamical symmetry breaking effects from lattice artifacts.  The next section discusses some ways this was done, but, while we believe the evidence is very good that the results presented here are not artifacts, it is not conclusive.  For this reason our results are labeled as ``indications.''

Dimensional analysis makes it apparent that the $F(B)$ should be defined in 
terms of an energy scale $T_0$, which can be identified with the characteristic 
scale of the microscopic degrees of freedom, or equivalently the ``microscopic 
temperature'' of the system.  Note that this tells us {\em only} about the 
density (and fluctuations of it) and is in general different from the mean 
free path of the interacting theory, which in the ideal hydrodynamic limit 
goes to zero.   Normally, for the Boltzmann equation (and, in AdS/CFT, classical gravity) to make sense, $T_0$ needs to be much smaller than the mean free path.  This work takes the opposite limit, since the mean free path is zero and $T_0$ is finite.

If Eq.~(\ref{lfluid}) is used to build a partition function, the 
``effective Planck's constant'' becomes dimensionful, as the ``microscopic 
gradient'' entirely factors out of the Lagrangian. 
Expansion around $T_0$, therefore, is potentially very different from the 
gradient expansion since ``non-perturbative'' contributions 
($\sim \exp[-1/(xT_0)]$, where $x$ is some distance scale) 
never quite go away.  Mathematically, this is analogous to 
non-Abelian gauge theory, where the action $\sim g_{YM}^{-2} F^2$ and 
likewise one cannot generally expand in powers of $g_{YM}$ except in some 
rigorously defined limits (the connection between non-perturbative physics in 
Yang-Mills theory and turbulence could in fact be more extended 
\cite{blaizot}; indeed, it has long been clear, and this paper may partially 
confirm, that Wilson loops and vortices share deep similarities).

It is straightforward to show that the classical expectation value of the 
energy-momentum tensor corresponding to the Lagrangian in Eq.~(\ref{lfluid}) 
is that of ideal hydrodynamics \cite{landau}
\begin{equation}
\label{ideal}
\ave{T_{\mu \nu}} = (p+\rho) u^\mu u^\nu + p g^{\mu \nu} 
\end{equation}
(using the ``mostly plus'' metric) and hence this is simply an unusual 
reparametrization of ideal hydrodynamics. 
The energy density and pressure in this notation are 
\begin{equation}
  \rho = -F(B) \label{eosrho} \eqcomma 
  p = F(B) - 2 B \frac{dF}{dB} \; . 
  \label{eosp}
\end{equation}
Hydrodynamic flow is defined as being perpendicular to any gradient of the 
$\phi^I$: $u^\mu \partial_\mu \phi^I = 0$. 
This, and $u_\mu u^\mu=-1$, unambiguously give 
\begin{equation}
  \label{flow}
  u^\mu = \frac{1}{6 \sqrt{B}} \epsilon^{\mu \alpha \beta \gamma}\epsilon_{IJK} 
  \partial_\alpha \phi^I \partial_\beta \phi^J \partial_\gamma \phi^K \; . 
\end{equation}
We can also show that $\partial_\mu (\sqrt{B} u^{\mu})=0$. 
By inspection, without any conserved charges (those are examined in
\cite{DerivExp}), one can identify 
\begin{equation}
  s = gT_0^3 \sqrt{B} 
\end{equation}
with the microscopic entropy. 
Using the Gibbs-Duhem relation, then, the temperature will be 
\begin{equation}
  T= \frac{\rho+p}{s} = T_0\frac{\sqrt{B}(dF/dB)}{g} \; . 
\end{equation}
Note the presence of $g$ as a free parameter.  This is the microscopic 
degeneracy, an intrinsic property of the system.  In the planar limit of 
Yang Mills theories, it tends to infinity faster than any other constant of 
the system, making the fluctuations discussed in this work irrelevant 
($T_0/g\rightarrow 0$ and any microscopic fluctuation gets distributed equally, by equipartition, between a ``large'' number of degrees of freedom).

The likely ubiquity of non-perturbative effects in this theory can be 
demonstrated by examining the ``vortex'' degrees of freedom, as was done in 
\cite{hydro1}:    
Naively, a vortex can be treated as a non-topological infinitesimal perturbation, a phonon $\pi_T$ equivalent to a soundwave $\pi_L$.
However, in a hydrostatic background, vortices {\em do not propagate}, 
yet carry arbitrarily small amounts of energy and momentum and interact in ways constrained by the symmetries of the Lagrangian ($\pi_L \pi_T \leftrightarrow \pi_L  \pi_L$ and $\pi_T \rightarrow \pi_L \pi_L$ are possible)
Thus, fluctuation-driven vortices can become stable, and their interactions 
could overwhelm the vacuum state. 
In \cite{hydrophoton} this interaction is demonstrated to have attractive 
components, leading to the likelihood of ``vortex condensates'' forming in the 
vacuum.  However, as \cite{hydrophoton} treats vortices as source terms, the 
equilibrium state of a liquid subject to fluctuations and its dependence on 
$T_0$ cannot be ascertained through such a perturbative expansion. 
The alternative analytical approach, to deform the theory in the infrared 
\cite{hydro0,gthydro}, is liable to give an incorrect vacuum since the 
deformation breaks the symmetries of the Lagrangian. 
This leaves the lattice as the only avenue to investigate vacuum properties 
consistently.

The theory formulated via Eq. \ref{lfluid} can be put on the lattice 
\cite{Creutz:1984mg,Rothe:1992nt} in the usual way, via 
\begin{equation}
\label{zdef}
\ln \mathcal{Z} = \int \mathcal{D}\phi_I \exp \left(i \int d^4 x L +J \phi_I
\right) \underbrace{\rightarrow}_{lattice+Wick} \int d \phi_I^i \exp \left[ 
  a^4 \sum_i  F(\phi_i) + J \phi_I \right] \; . 
\end{equation}
Throughout this work we use the ideal Bose gas EoS, 
\begin{equation}
  \label{Fdef}
  F(B) = - T_0^4 B^{2/3} \; . 
\end{equation}
This can be easily generalized to any monotonic EoS  (without phase 
transitions), for example an EoS, say, fitting the QCD cross-over 
\cite{gthydro}. 
The ideal gas, however, is a good testing laboratory as all of its parameters 
are very simple 
\begin{equation}
\label{expect}
 \ave{\rho} = \frac{g^4\pi^2}{30} T^4 \eqcomma \ave{p}=\frac{\ave{\rho}}{3} 
 \eqcomma \ave{s} = \frac{2g^3\pi^2}{45} T^3   . 
\end{equation}

Again, note that the starting point in this formalism requires the 
identification of $s \propto \sqrt{B}$ 
and, from the formulas above, the following relation: 
\begin{equation}
\label{expect2}
 \ave{\rho} = \frac{\pi^2}{30} \left( \frac{45}{2\pi^2} \right)^{4/3} 
 \ave{s}^{4/3}  , 
\end{equation}
where the constant of proportionality ($\approx 0.987$) can be absorbed 
into the ratio of scales $(aT_0)^4$, giving the above $F(B)$.

\subsection{Lattice implementation}

Since centered differences lead to two disjoint, interwoven lattices, our 
lattice derivatives are approximated by an average of all (eight) one-sided 
differences per hypercube. 
All quantities derived from the field derivatives $\partial_\mu\phi^I$ (e.g., 
$B_{IJ}$, $u_\mu$, $T_{\mu\nu}$) are therefore situated at the centers of the 
hypercubes. 
To handle the periodic nature of the toroidal lattice, we subtract the 
hydrostatic background (when $\phi^I=x^I$) and deal with the ``shifted'' 
coordinates: 
\begin{equation}
\label{pidef}
  \pi^I = \phi^I - x^I \;\; \to \;\; 
  \partial_\alpha\phi^I = \partial_\alpha\pi^I + 1 \delta_\alpha^I \; .
\end{equation}
The $\pi^I$ are the fluid phonon fields.

Because we expect extended structures (e.g., vortices) to arise, we use HMC 
updates. 
The required variation of the action with respect to local field values is 
given by: 
\begin{eqnarray}
  \frac{\delta S}{\delta\phi^I(x)} = 
  \frac{\delta S}{\delta \sqrt{B}} \frac{\delta \sqrt{B}}{\delta(\partial_\alpha\phi^J)} 
  \frac{\delta(\partial_\alpha\phi^J)}{\delta\phi^I(x)} 
\end{eqnarray}
\[\ = \sum_{y,\mu,\nu,\sigma} \frac{dF}{d\sqrt{B}} \delta^{IJ} 
  \delta(y-x\pm\hat\mu/2\pm\hat\nu/2\pm\hat\sigma/2) 
  \left. \frac{\sqrt{B}}{8} B_{JK}^{-1} \, |\epsilon_{\mu\nu\sigma\alpha}| \, 
  \partial_\alpha\phi^K \right|^{y - \hat\alpha/2}_{y + \hat\alpha/2} \; . 
\]
The updating algorithm and the calculation of observables has been implemented 
in C code (including the ranlux random number generator 
\cite{Luscher:1993dy}), with multicore parallelization via OpenMP.

Interpreting the lattice data is complicated by the fact that this is a 
manifestly non-renormalizable theory: The continuum limit is one where 
$T_0 \rightarrow \infty$, and the approach to it is obviously divergent 
for any dimensionful quantity, though not for dimensionless numbers.

This however is physically reasonable, since the regulator in this theory can 
be interpreted as the approach to the physical microscopic scale.  Varying 
$aT_0$, the lattice spacing in units of the microscopic scale, we are 
considering different systems whose ratio of the characteristic size of 
microscopic excitations to collective excitations is different.  This, in 
this context, is the only free parameter of the theory. 
(However, another possible one arises via the ``macroscopic'' temperature, 
the period in the time direction of the finite temperature lattice, and 
the influence of this should be investigated as well.)

The continuum limit would be an infinite lattice of a given $aT_0$. 
This limit is obviously not reachable, or even approachable by the finite 
computing resources used so far in this project. 
Our results thus far, however, show that we may be closing in on such a limit.

\section{Results}
\subsection{Evidence for coexisting phases \label{entropyres}}

\begin{figure}[h]
\begin{center}
\includegraphics*[clip,width=7.5cm]{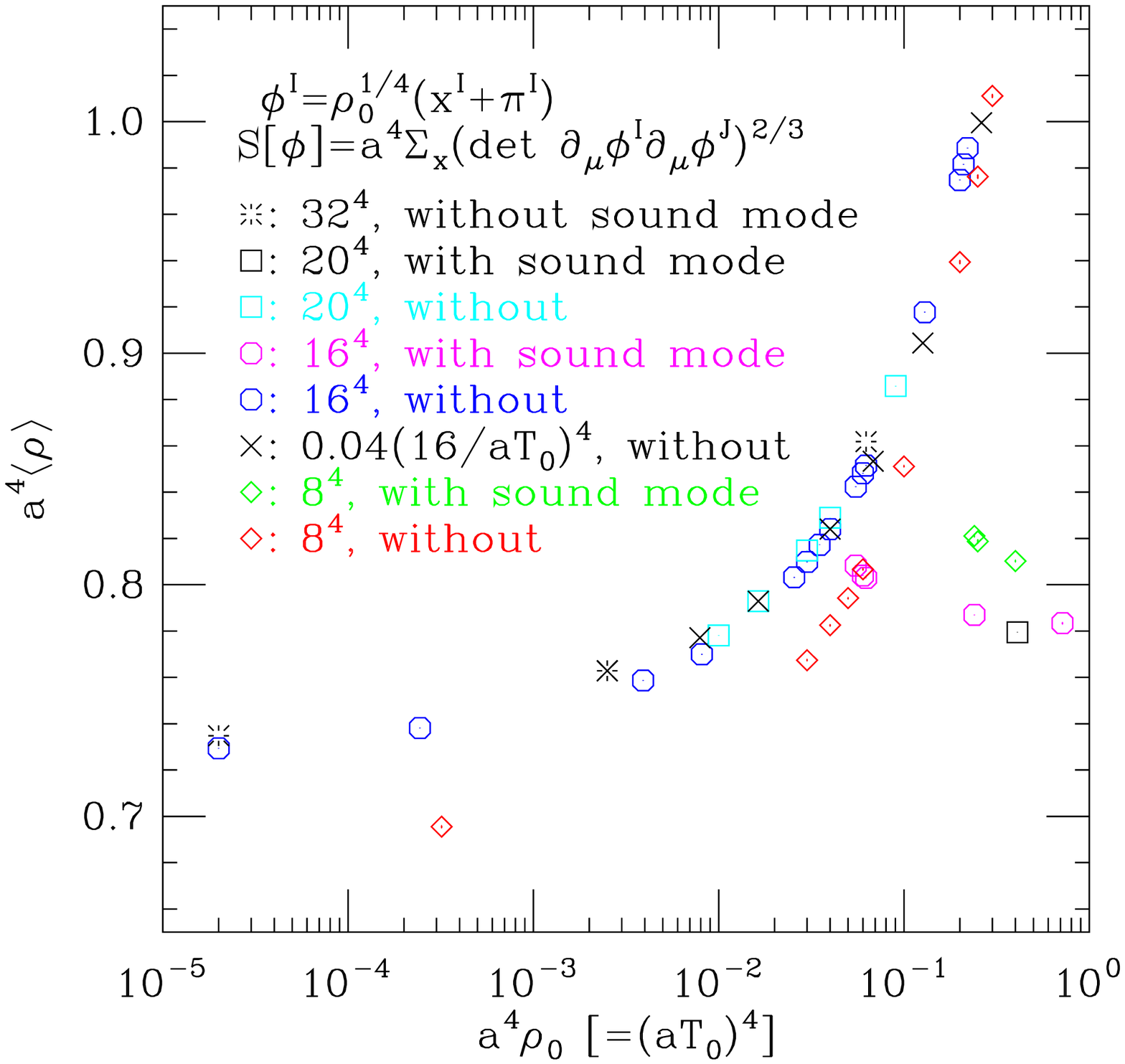}
\includegraphics*[clip,width=7.5cm]{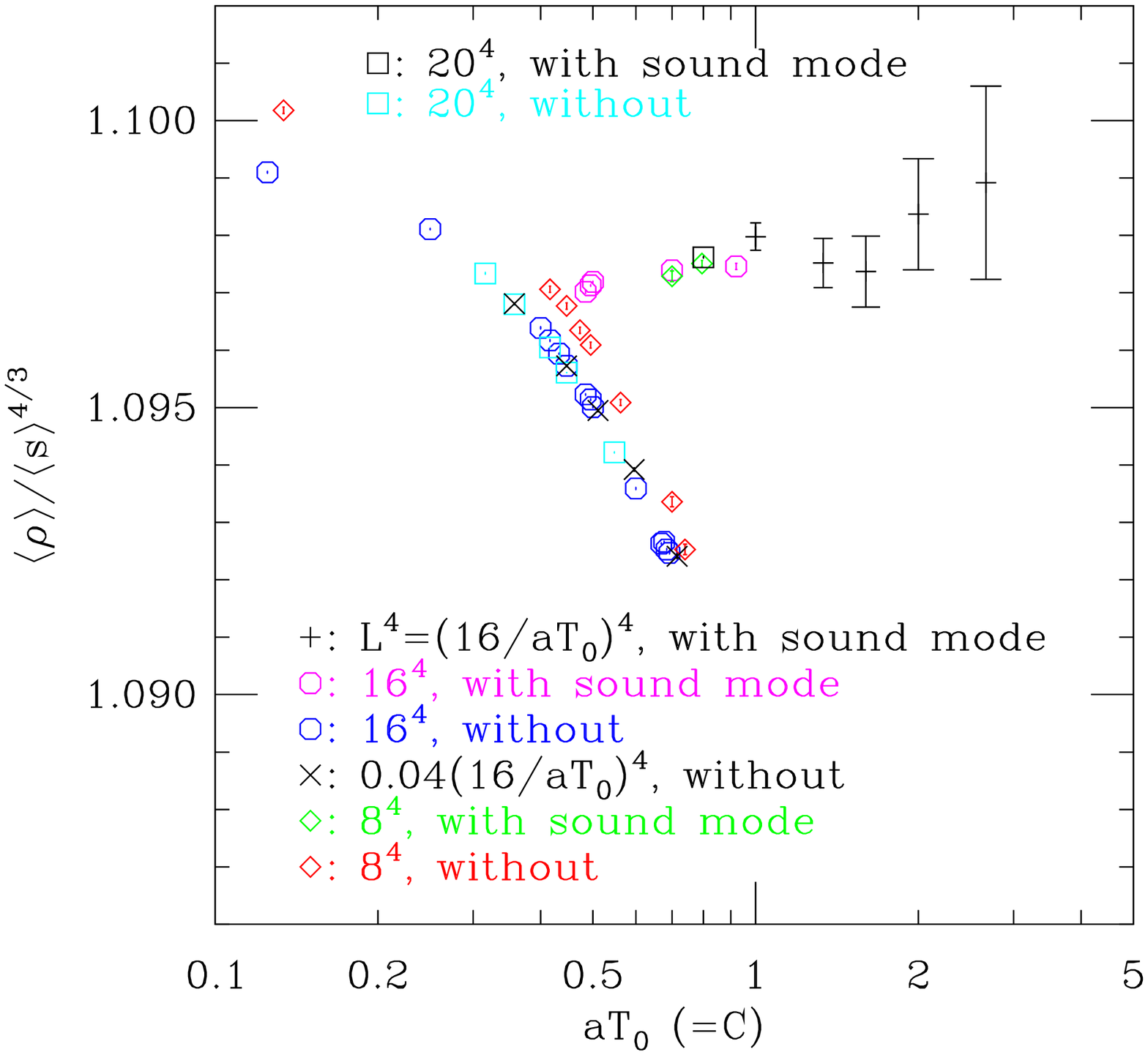}
\caption{\label{figs} The energy density as a function of $aT_0$ and the 
  lattice size for ensembles both with and without collective ``sound modes'' 
  present. Left: Average energy density in lattice units. Right: ratio of 
  average energy density and the $4/3$-power of the lattice-averaged entropy 
  density, $\langle s\rangle$.}
\end{center}
\end{figure}

The overall energy density $\rho$ and its relation to entropy density $s$ are explored in Fig. \ref{figs}.

Beyond a ``critical'' $aT_0$ ($\approx 0.47$ for the $16^4$ lattices), 
two vacua are clearly visible. 
The upper vacuum's entropy and energy has a monotonic power law dependence on 
$aT_0$, and can be readily identified with a state where the bulk of the 
entropy is carried by microscopic degrees of freedom. 
The lower vacuum has a nearly constant microscopic entropy with respect to 
$aT_0$, with a nearly constant ratio of energy to entropy density, and can be 
identified with the vacuum where entropy and energy density has gone 
into collective (``sound'') modes. 
One can see in the plot on the right that corresponding entropy-density 
correlations have managed to maintain an enhancement in 
$\ave{\rho}/\ave{s}^{4/3}$ when such modes are present. 

To interpret this data we note that ``low'' and ``high'' entropy states refer 
only to the microscopic entropy, since we have no way, at present, of 
measuring the entropy contained in the macroscopic perturbations 
(see, however, the Discussion section, specifically around Eq. \ref{intermeasure}). 
Thus, the enhancement of $\ave{\rho}/\ave{s}^{4/3}$ is in line with the 
hypothesis that macroscopic degrees of freedom in the ``low entropy'' 
phase carry a significant fraction of entropy, as well as energy density.

The locations of the vacua show a small dependence on the lattice size, 
indicated by the different symbols in the figures, with the jump to the 
$20^4$ and $32^4$ lattices thus far showing little difference from the $16^4$.

Transitions between vacua have also been observed, but appear to be exceedingly 
rare. 
Close to the ``critical'' point, we have observed single transitions from the 
lower entropy state to the higher one; and well above this point, we see 
single tranisitons from the higher entropy state to the lower one. 
Hence, thus far we are not able to verify the scaling of these transitions with 
system volume, which, if its rate $\Gamma$ went as 
\begin{equation}
\label{scalvol}
\Gamma_{1 \leftrightarrow 2} \sim \exp\left[  - \left| s_1 - s_2 \right| V 
  \right] 
\end{equation}
would be a clear indication of first-order behavior.

\begin{figure}[h]
\begin{center}
\includegraphics*[clip,width=7.5cm]{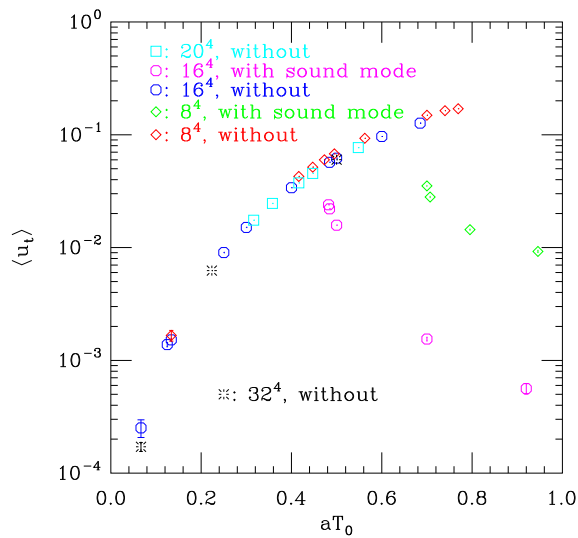}
\caption{\label{ut} Plot of the time component of the flow as a function of $aT_0$.}
\end{center}
\end{figure}

Figure \ref{ut} compounds these indications by examining the average of 
the (Euclidean) time component of the flow $u_0$. 
While spacelike components ($u_i$) are on average zero if rotational symmetry is not broken (and 
by observation it is not), $\ave{u_0}$ shows a clear non-zero value rising steadily 
towards the hydrostatic limit ($\to 1$). 
Along the way, however, the lower-entropy vacuum arises and transitions to it 
result in substantially increased spatial velocity fluctuations, lowering 
$\ave{u_0}$.

This flow is {\em not} fully explainable in terms of perturbations, as can be 
seen by the fact that, for higher values of $aT_0$, the average value of the 
flow is multi-valued, something only possible in the presence of a first order 
transition:  In this regime, the vacuum where $u_0$ is lowest (and hence
where fluctuations in $u_i$ are higher) has an average expectation value of 
collective excitations, which agrees well with the idea that this vacuum is 
dominated by fluctuation-driven turbulence.   

In the future, correlators of flow observables such as vorticity 
\begin{equation}
  \label{vortex}
  C_{P} = \oint_P (p+e) u_{\mu} dx^{\mu} 
\end{equation}
or of the tensor shear 
\begin{equation}
  \label{omega}
  \ave{\Omega_{\mu \nu}} =\ave{ u_\mu u_\nu + g_{\mu \nu}} = \ave{ B_{IJ}^{-1} 
    \partial_\mu \phi^{I}  \partial_{\nu} \phi^J} \; , 
\end{equation}
(the average values are fixed to zero by symmetry) can be examined 
in the two regimes to see what difference is there between the flow 
structures of the two vacua.

\subsection{Evidence for ``sound'' modes}

\begin{figure}[h]
\begin{center}
\includegraphics*[clip,width=7.5cm]{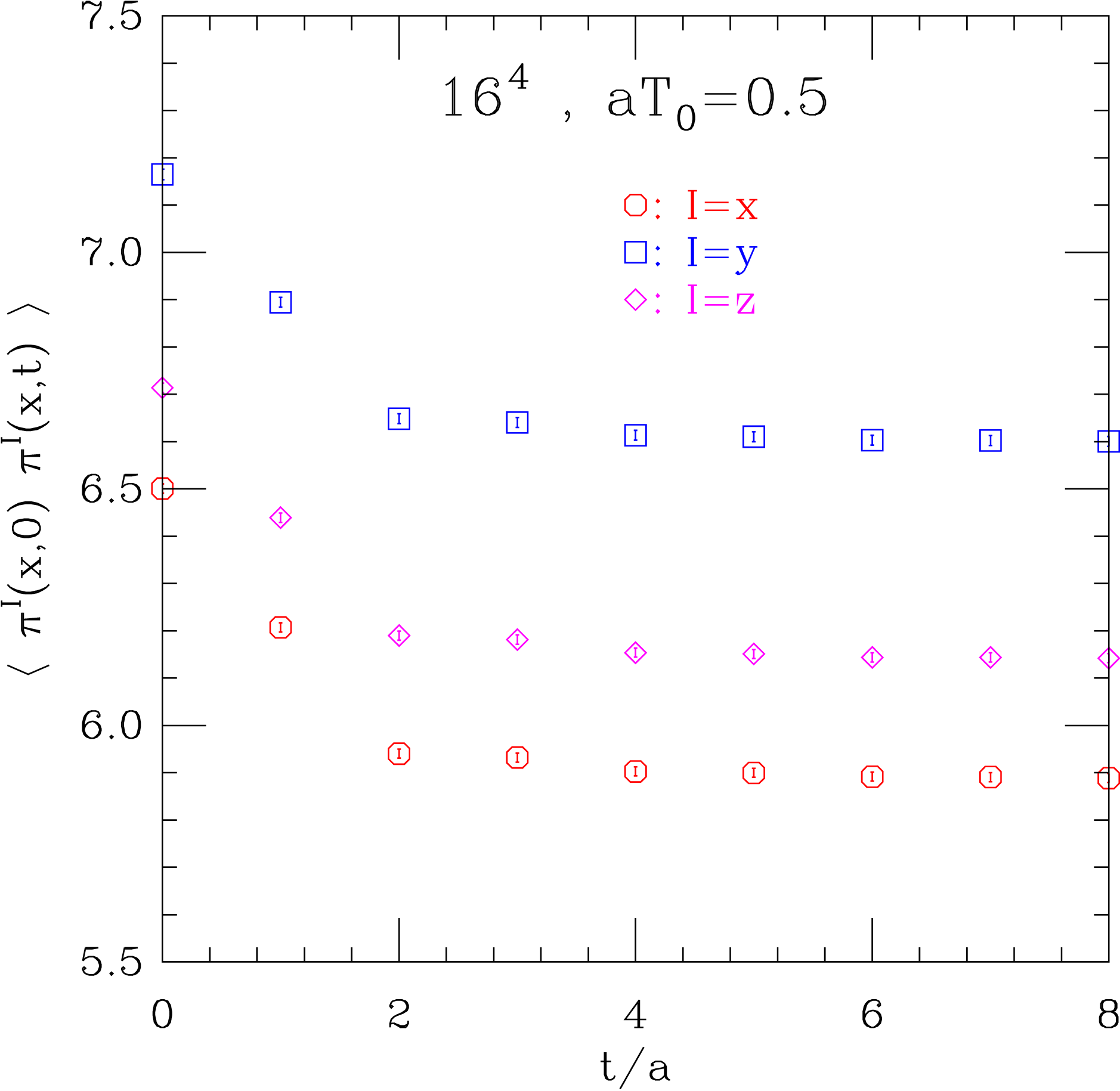}
\includegraphics*[clip,width=7.5cm]{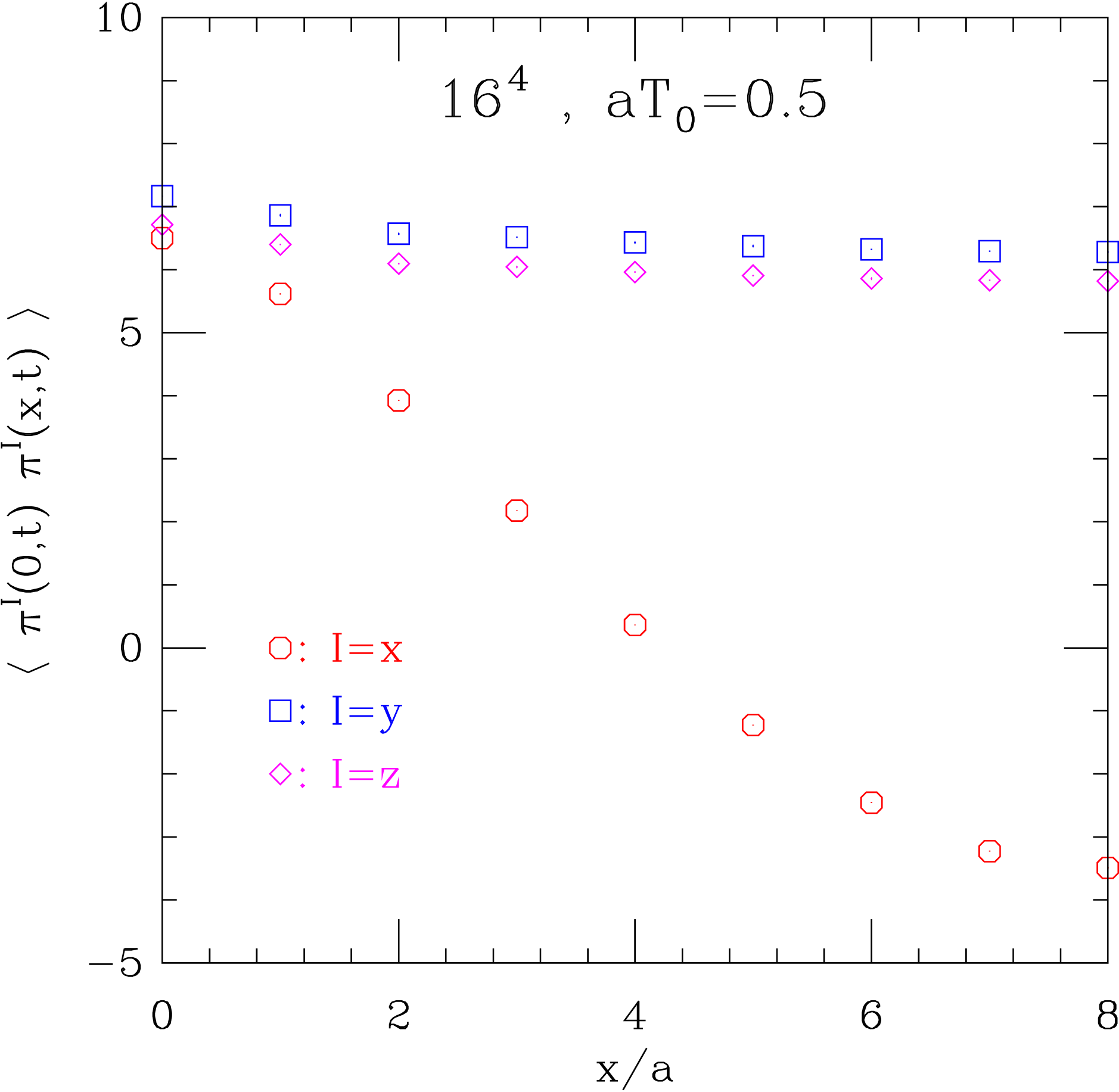}
\caption{\label{modefig} Phonon correlators of longitudinal ``sound'' modes.}
\end{center}
\end{figure}

Figure \ref{modefig} shows the space-time dependence of the (hypercube 
averaged) phonon correlators. 
On this ensemble, one can clearly see a nearly time-independent longitudinal 
mode in $\langle \pi^x(0) \pi^x(x) \rangle$. 
The absolute magnitude of the $\pi^I$ fields does not matter, but such 
correlations will clearly enter into the derivatives $\partial_\mu\pi^I$ 
(and hence $u_\mu$, etc.). 
``Transverse'' modes (e.g., sinusoidal $x$-dependence in 
$\langle \pi^y(0) \pi^y(x) \rangle$) are also sometimes seen, sometimes 
in combination with the longitudinal ones.

\begin{figure}[h]
\begin{center}
\includegraphics*[clip,width=7.5cm]{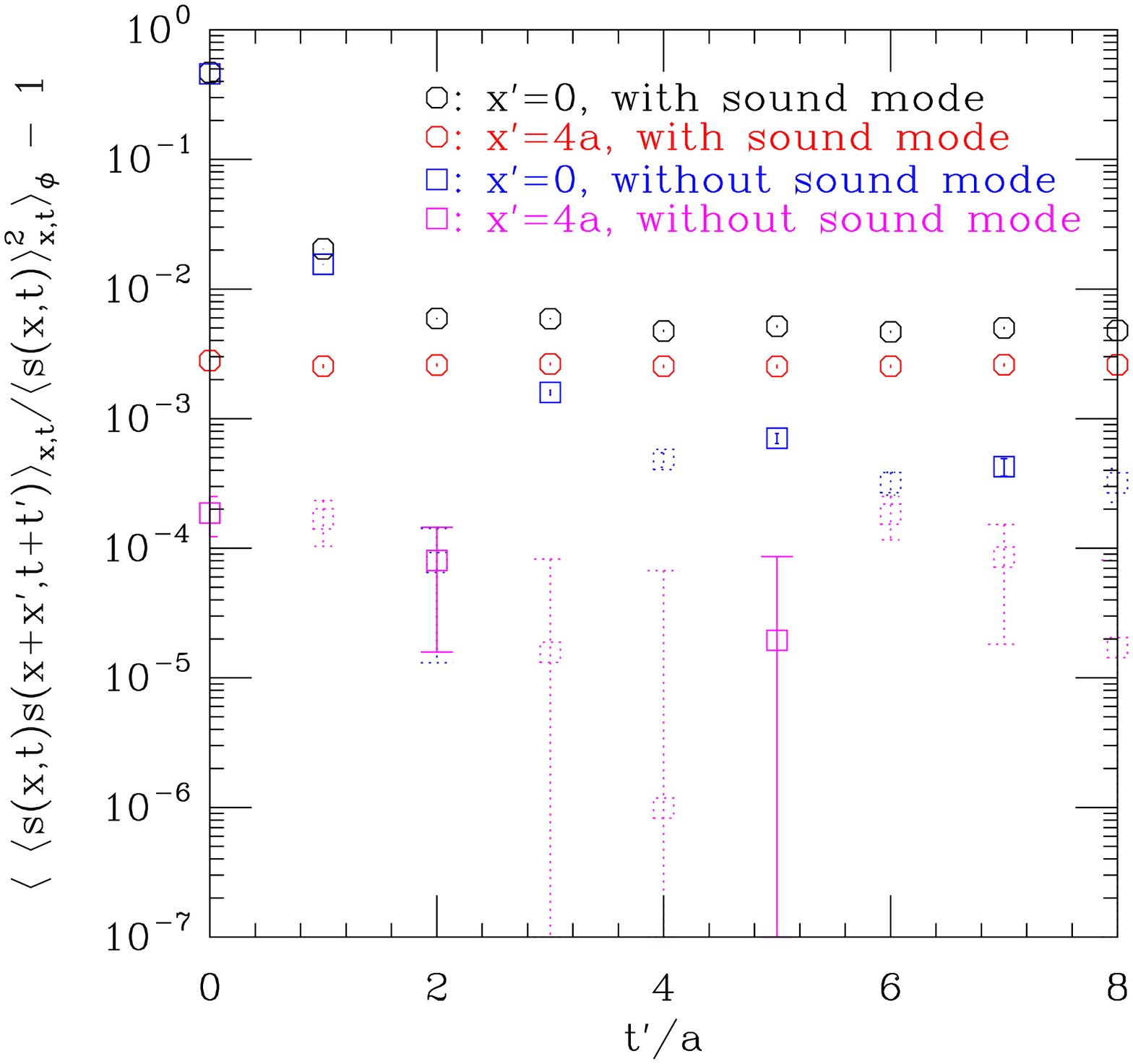}
\includegraphics*[clip,width=7.5cm]{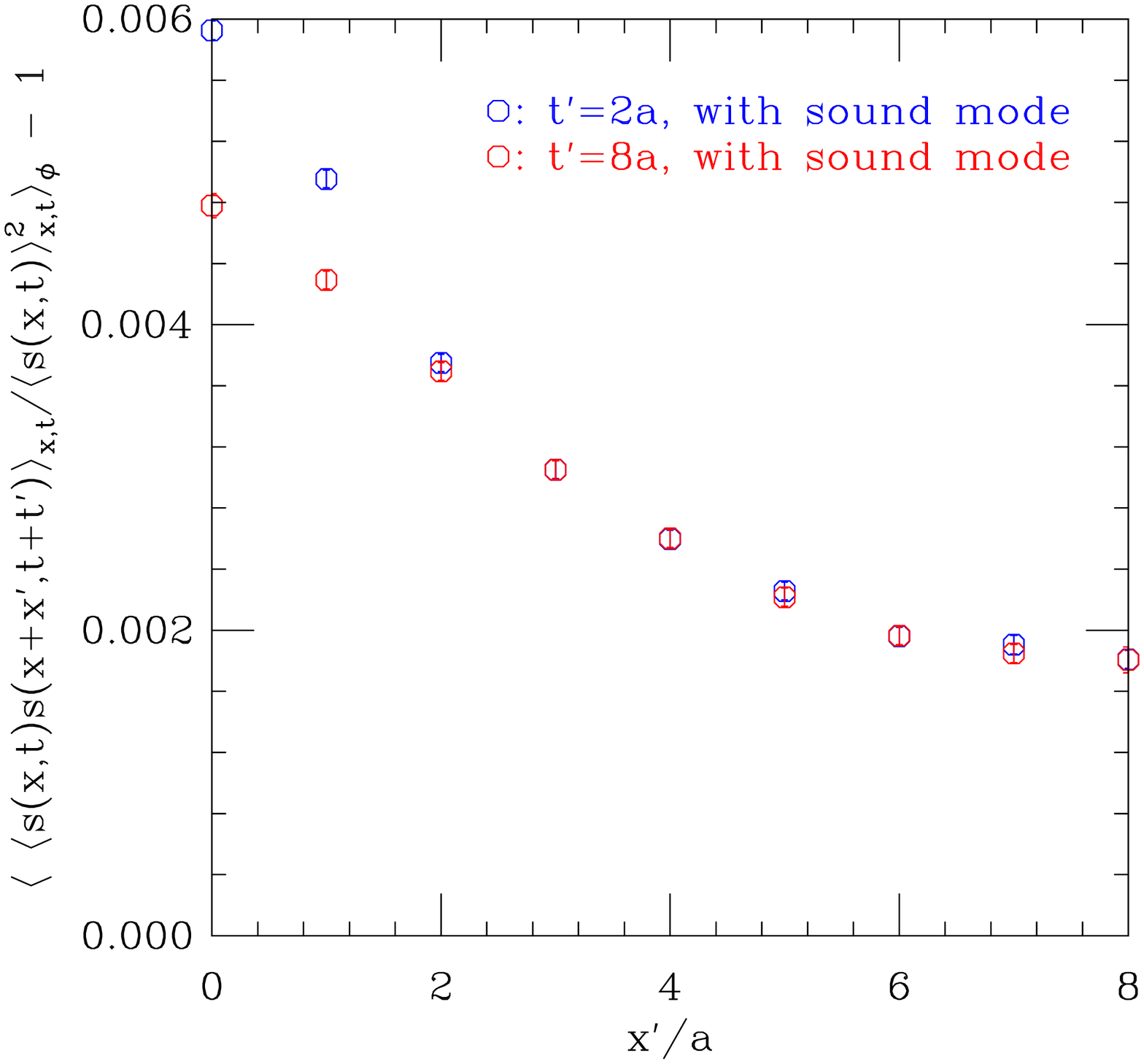}
\caption{\label{ssfig} Relative entropy-density correlators. 
Left: results for two $16^4$, $aT_0=0.5$ ensembles -- one with a ``sound 
mode'' and one without -- as a function of time separation (dotted symbols 
are for negative values). 
Right: the spatial dependence from the sound-mode ensemble.}
\end{center}
\end{figure}

In Fig.~\ref{ssfig} we plot corresponding relative entropy-density 
correlators. 
Looking at the plot on the left, one can see that such correlations 
fall off into noise rather fast for ensembles without sound modes 
(i.e., from the higher energy branch), whereas for the lower energy 
ensemble, they persist across the lattice. 
It is straightforward to relate these persistent sound modes to the 
average value of the flow component in Fig.~\ref{ut}. 
However, as the next subsection explains in detail, in a finite 
system, care must be taken to interpret these degrees of freedom as 
mere sound perturbations.

\subsubsection{Interpretation in terms of kinks and domain walls?}

Another feature of first-order phase transitions is the presence of 
kink solutions in the mixed phase:  If two global vacua are possible, 
a translationally dependent interpolation between them is also a minimum. 
Such objects, in conventional phase-transition physics, are called domain 
walls, solitons, instantons (if time-dependent), calorons (if at finite 
temperature), and so on.

In the context of classical hydrodynamics, a natural implementation of 
this concept is the relativistic static shock: long-lived shocks, while 
generally unstable against small perturbations, arise as classical 
solutions to the hydrodynamic equations of motion. 
Relativistically, the Taub adiabat is the most well-known example 
\cite{taub,landau,zeldovich}. 
If the equation of state is not single-valued, these shocks can be 
stationary \cite{landau,zeldovich} and, in presence of a surface tension, 
stable.  Such topological objects form the surface of objects such as 
bubbles, which characterize the mixed phase of a first-order phase 
transition. In a sudden cooling, they can also remain in the ``colder'' 
phase, where symmetries of the hot phase are broken.

In the context of our hypothesized first-order phase transition between the 
turbulent and the hydrostatic vacuum, the kink should appear as a long-lived 
shock between a hydrostatic and a boundary layer. 
The shock's width could be interpreted as a ``lower limit'' to Prandl's 
boundary layer \cite{prandtl}, which would reinforce our contention that 
quantum hydrodynamics can be seen as a lower limit of viscosity.

Indeed, Fig.~\ref{modefig} shows a correlator of the flow field $\pi^I$ 
(See Eq.~\ref{pidef}) where a space-like mode, stable in time, can be seen. 
The stationary nature of this mode precludes it from being a perturbative 
expansion, but it is somewhat expected if the dynamics of the vacuum 
is driven by vortices, since vortices are stationary \cite{hydro1}.

Its space dependence could be seen as an indication of a topological 
structure, but for conclusive evidence of this we would need to study 
{\em three} point functions of correlators such as Eqs.~(\ref{vortex}) 
and (\ref{omega}), to confirm that the vacuum structure around a kink 
interpolates between the two phases.

\section{Discussion}

\subsection{Effects of the lattice formulation}


As mentioned earlier, we cannot remove the finite lattice spacing, but we 
can alter its effect by choosing different discretizations for the lattice 
derivatives. 
More extended finite differences are certainly possible and we could even 
``take a step backward'' and move to centered differences to explore the 
effect of finite $a$. 
(Much older results \cite{TBtalk}, from the start of this project, using 
centered differences, indicate a ``critical'' point around $aT_0 \approx 0.7$ 
on $8^4$ lattices, close to where we see one now on lattices of the same size.)


Since the collective modes which we see on our lattices tend to fit with one 
wavelength within the finite, periodic box (see Fig.~\ref{modefig}), we cannot 
help but wonder whether these are simply symptoms of the toroidal nature of 
the volumes they occupy. 
It may be worthwhile switching to Dirichlet ($\pi^I=0$) or Neumann 
($\partial_I\pi^I=0$) boundary conditions to see how they behave there. 
But we point out that these modes also arise from the particular action 
in question (see Eq.~\ref{Fdef}) and that they more readily appear for larger 
volumes, making us believe that they (and their effects) will persist in any 
``thermodynamic limit'' ($V \to \infty$).


Another consequence of using the lattice formulation is that we necessarily 
have a Euclidean signature (see, e.g., $\ave{u_0}$ in Fig.~\ref{ut}). 
One might imagine that, close to the hydrostatic limit 
($\ave{u_0} \lesssim 1$), we could map our Euclidean hydro to a Minkowski 
one: e.g, 
\begin{equation}
  u_\mu = ( \pm\sqrt{1-v^2} , \vec v ) = ( \pm1/\gamma , \vec v ) \;\; \to 
  \;\; u_\mu = ( -\gamma , \gamma \vec v ) \; , 
\end{equation}
but we find that we are still quite far from this limit when the collective 
modes (and larger velocity fluctuations) appear, driving the limit beyond our 
control.

Before moving on we make one more observation about the two ``phases'' we 
observe: larger spatial volumes drive the higher energy state to even higher 
energy and entropy and the lower energy phase to lower energies, but 
larger time extents have the opposite effect for only the lower-energy 
(collective-mode) phase. 
We may be in a position to explore the ``macroscopic'' thermodynamics of these 
collective excitations. 
The interaction measure ($I=\rho-3p$) can expressed as 
\begin{eqnarray}
\label{intermeasure}
  I_{macro} + I_0 &=& - \frac{T_{macro}}{V} \; \frac{d \ln Z}{d \ln a}  \nonumber \\
  &=& - \frac{T_{macro}}{V} \; 4 \left( 1 + \frac{d \ln T_0}{d \ln a} \right) 
  \langle S[\phi^I] \rangle  \nonumber \\
  &=& - 4 \left( 1 + \frac{d \ln T_0}{d \ln a} \right) \langle \rho \rangle , 
\end{eqnarray}
where we used $T_{macro}=1/(aN_t)$ and $V=(aN_x)^3$, and $I_0$ represents the 
zero-temperature (divergent) part which must be subtracted off. 
So we see the interaction measure of collective excitations is related 
directly to the average microscopic energy density. 
From here, the integration method (see, e.g., \cite{IntEoS}) can be used to 
arrive at the macroscopic pressure, energy and entropy density.

\subsection{Further developments}

The main barrier to our results not being more conclusive are the 
computational limits.   This paper is invariably a series of several 
ones, with higher statistics runs giving a more coherent picture of 
the situation.

The availability of more statistics will allow us to understand if 
the transition between the two vacua follows the volume scaling law 
expected from a first-order phase transition (i.e., Eq.~\ref{scalvol}). 
We would regard this as definite proof hydrodynamic fluctuations 
trigger a first-order phase transition in the vanishing viscosity 
limit. This proof is at the moment lacking.

In Section \ref{entropyres} we have proposed investigating turbulence 
using vorticity and flow correlators.  Such correlators are well-studied in 
classical turbulence \cite{zakharov}.  In particular, if our vacuum is similar 
to classical hydrodynamic turbulence, we would expect that they fall as a 
power-law, with an exponent related to turbulent spectral exponents 
(such as the well-known Kolmogorov spectrum) calculated in semiclassical 
theories.  It would be very interesting to investigate this further.   
 Studies such as these are instrumental
in formulating the effective Lagrangian of the colder phase.  If our interpretation is correct, turbulence results in the symmetries of hydrodynamics being broken.   Thus, the effective Lagrangian of the cold phase will look like
\begin{equation}
L_{eff} = F(B) + \sum_i a_i(T_0) f(B_{II})  + \sum_i b_i (T_0) g(B_{IJ}) + \sum_i c_i h(\phi_I)
\end{equation}
where coupling constants $a_i,b_i,c_i$ break the conformal, isotropy, and homogeneity conditions respectively.   For a true dynamical symmetry breaking effect rather than a lattice artifact, these terms must arise as a phase transition rather than a smooth cross-over, something which we seem to observe.
Furthermore, operators of dimension $D$ should be $\ll T_0^D $ if the theory examined here is indeed an effective theory valid at scales much smaller in momentum space than $T_0$.

This study could lead to a full non-perturbative renormalization group 
analysis of hydrodynamics as a field theory.   Naively, this field theory is 
non-renormalizeable, something seen both on dimensional grounds (appearance of 
operators of dimension higher than four) and on physical grounds (the 
coarse-graining scale in hydrodynamics is a physically observable quantity). 
This question is directly connected to the question of whether ``ideal 
hydrodynamics'' with the coarse-graining scale going to infinity (in other 
words the limit $\eta/s \rightarrow 0$) is well-defined.   In general, for an 
observable $X$, iff $\lim_{\delta\rightarrow 0} \ave{X} \sim \ave{X_0}_{termostatic}$, 
the thermostatic state is stable. 
If $\lim_{\delta \rightarrow 0} \ave{X}/\ave{X_0} \sim f(B)$, the anomalous 
dimensions, caused by quantum turbulence, will affect this operator. 
If $\lim_{\delta\rightarrow 0} \ave{X}/\ave{X_0} \sim \delta^{-\alpha}$ or 
$\sim \exp(\alpha \delta^{-1})$ for some $\alpha$, the theory is 
``renormalizeable'' for some observables, while if 
$\lim_{\delta\rightarrow 0} \ave{X}/\ave{X_0} \sim \delta^{-\alpha}$ or 
$\sim \exp(\alpha \delta^{-1})$ for $\alpha$s that are $\ave{X}$-specific (one 
$\alpha$ for the scalar and another for the tensor, defined below) the theory 
is ``trivial'', in that taking $\delta \rightarrow 0$ makes the vacuum 
diverge. 
In the latter case a limit $\eta/s \rightarrow 0$ is indeed inconsistent, and 
the degree of divergence could be used to understand the behavior of the 
limiting $\eta/s$ on the EoS. 
The statistics required for a systematic study of important $\ave{X}$ is at 
the moment well beyond our capabilities, but in principle this study is 
achievable with today's computing technology.

Scaling in the number of dimensions is also interesting: 
In 3D, vortices can point in random directions, and tend to form an 
instantaneously disordered ``glass'' which, over a long period of time, seeds 
the well-known Kolmogorov cascade \cite{cascade}. 
In 2D, vortices can only point in two directions.  Given the generally 
attractive potential between vortices, the turbulent phase in 2D is 
characterized by a 2D crystal of more or less ``regular'' structures 
\cite{coherent}, which over long periods of time quenches smaller structures 
into larger ones, thus motivating an ``inverse cascade'' picture (and 
effectively negative transport coefficents \cite{rom2d}).  In 1D, vortices 
are absent altogether, with non-linear corrections being given exclusively by 
sound-waves. As sound waves, unlike vortices, undergo normal scattering 
\cite{hydro1}, 1D turbulence could be described by the kinetic approaches 
explored thoroughly in \cite{zakharov}. If the phase we found can be 
characterized as fluctuations-induced turbulence, dimensional scaling 
will have the properties outlined here.

Given a larger lattice, the structure of the solitonic modes can also be 
analyzed in space, to extract the Prandl width and see how the two phases, if 
this is indeed what they are, vary in observables such as Eqs.~(\ref{vortex}) 
and (\ref{omega}).  Eventually, given the rich pure gauge lattice QCD data in 
this area, this opens the door to quantitative exploration of the analogies 
conjectured in \cite{blaizot}.

If our picture of a non-trivial ``turbulent'' vacuum is confirmed, it is still 
completely unclear to what extent this applies for $\eta/s \ne 0$.  For a finite 
viscosity, the partition function of Eq.~(\ref{zdef}) will aquire imaginary 
components, representing the dissipation of collective degrees of freedom into 
microscopic ones.  Lattice techniques described in this work cannot be applied 
to this system, and it is not clear to what extent the extended collective 
excitations are fragile against decohering degrees of freedom.  Hence, we still 
do not know in what regime the gradient expansion would fail for the reasons 
discussed here.  The dimensional analysis arguments in terms of sound waves 
given in the Introduction could provide a good benchmark, but this is not a 
rigorous proof. 
Nevertheless, the effort here can be taken as the first step to define an 
effective field theory of hydrodynamics that includes {\em both} 
coarse-graining parameters: the mean free path and the fluctuation scale. 
Perhaps a mixed classical and quantum effective theory \cite{ctp} could 
provide more quantitative answers.

In conclusion, we found intriguing hints of a non-trivial vacuum structure in 
ideal hydrodynamics without chemical potentials.  If confirmed, it could 
potentially invalidate the gradient expansion as a complete effective theory 
for a range of microscopic parameters of the underlying theory.  We hope and 
expect subsequent work will clarify the properties of the non-trivial phase 
and the transition between the two.

GT acknowledges support from FAPESP proc. 2014/13120-7 and CNPQ bolsa de 
produtividade 301996/2014-8. 
Simulations were performed at the Institute for Theoretical Physics at 
Uni-Regensburg and we thank the Sch\"afer and Braun Chairs for continued 
access and the administrators for smooth machine operation. 
TB acknowledges his family's patience and support of what has become his 
late-night hobby. 
Both authors acknowledge the old LSS group at the University of Arizona for 
insightful discussions that eventually led to this work.

\end{document}